# A MULTI-FACTOR SECURITY PROTOCOL FOR WIRELESS PAYMENT- SECURE WEB AUTHENTICATION USING MOBILE DEVICES

Ayu Tiwari, Sudip Sanyal
*Indian Institute of Information Technology (IIIT), Allahabad (UP)*
*ayu.tiwari@wcc.iiita.ac.in, ssanyal@iiita.ac.in*

Ajith Abraham, Svein Johan Knapskog
*Centre for Quantifiable Quality of Service in Communication Systems, Norwegian University of Science and Technology, Trondheim, Norway, {ajith,knapskog}@q2s.ntnu.no*

Sugata Sanyal
*School of Technology & Computer Science, Tata Institute of Fundamental Research, (TIFR), Mumbai*
*sanyal@tifr.res.in*

**ABSTRACT**

Previous Web access authentication systems often use either the Web or the Mobile channel individually to confirm the claimed identity of the remote user. This paper proposes a new protocol using multifactor authentication system that is both secure and highly usable. It uses a novel approach based on Transaction Identification Code and SMS to enforce extra security level with the traditional Login/password system. The system provides a highly secure environment that is simple to use and deploy, that does not require any change in infrastructure or protocol of wireless networks. This Protocol for Wireless Payment is extended to provide two way authentications.

## 1. INTRODUCTION

As computing becomes pervasive, people increasingly rely on public computers to do business over the Internet. Now, the Internet has become the preferred environment for a multitude of e-services: e-commerce, e-banking, e-voting, e-government, etc. Security for these applications is an important enabler. Financial institutions offering Internet-based products and services to their customers should use effective methods to authenticate the identity of customers. Accessing today's web-based services invariably requires typing a username and password to authenticate. This is a significant vulnerability since the password can be captured by the public computer and later reused by a hostile party. The current online payment systems do not provide adequate user authentication. Thus it is possible for an unscrupulous user to enter credit card number or account details stolen from valid users. Financial agencies consider single-factor authentication to be inadequate for high-risk transactions involving access to customer information or the movement of funds to other parties using web browsers or cell phones/PDA's. So we need Multifactor Authentication techniques to secure our web transactions and to increase faith of users on mobile financial transactions. In this paper we propose an authentication system that is both secure and highly usable, based on multifactor authentication approach. It uses a novel approach to create an authentication system based on TIC's (Transaction Identification code) and SMS to enforce an extra security level for the traditional login in a username/password context. Al-Qayedi et al. [1] have also proposed the use of SMS but have not used TIC's in their protocol. TIC's are user specific unique transaction identification codes which are issued by banks or financial institutions to the user. This code is similar to One Time Password (OTP) and a code is used only once. This paper also suggests an encryption/decryption technique that would be used to keep TIC's as secret codes on cell phones/PDA's. The user can easily pick up a TIC (from the stored list of TIC's) to initiate secure web transaction using cell phones/PDA's, instead of typing a complicated TIC code in each transaction. This paper also describes a system for two way authentication i.e. the company or service provider is authenticated to the user along with the authentication of user to the financial institution.

The paper is structured as follows: Section two reviews the related work on e-payment systems. Section three introduces the Multifactor authentication approach. Section four presents our protocol for Wireless Payment including the system design and architecture for secure web authentication. Section five presents the architecture of the two way authentication scheme and its functional components. In Section six, we discuss some implementation issues followed by some conclusion and some ideas for future work in Section seven.

## 2. BACKGROUND AND RELATED WORK

According to Gao et al. [2], mobile payment refers to wireless-based electronic payment for m-commerce to support point-of-sale/point-of-service (POS) payment transactions using mobile devices. In general, m-payment systems can be used by wireless-based merchants, content vendors and information and service providers to process and support payment transactions driven by wireless-based commerce applications. As discussed in [2], the existing m-payment systems can be classified into two major types. The first type is account-based payment systems which can be mobile phone-based, smart card or credit-card m-payment systems [3, 4, 5, 6]. The second type of m-payment system refers to mobile POS payment systems that enable customers to purchase products on vending machines or at retail stores with their mobile devices. This payment system is designed to complement existing credit and debit card systems for mobile users and can be either automated POS payments or attended POS payments. An example of mobile POS payment system is Ultra's M-Pay (http://www.ultra.si/).

## 2.1 SECURE ELECTRONIC TRANSACTION (SET)

The Secure Electronic Transaction is an open encryption and security specification designed to protect credit card transactions on the Internet. The companies that collaborated in the development of SET include IBM, Microsoft, Netscape, RSA, Terisa and Verisign. It is supported by major corporations such as VISA Inc. and MasterCard. Although SET have been designed to operate in a wired infrastructure [7, 8, 17], its transaction flow and implementation of security are of interest to us since it can also be employed in a wireless scenario. The SET protocol is an evolution of the existing credit-card based payment system and provides enhanced security for information transfer as well as authentication of transaction participant identities by registration and certification. SET is also an international standard with published protocol specifications. While SET permits customers to make credit-card payment to any merchant offering web-based services, customers also have the option of paying for other types of services using the on-line banking facilities. A brief description of SET is given in Figure 1.
1. The consumer accesses the merchant's web site, browses the goods on display and selects what he or she wants and gets the total cost of all chosen items including taxes and shipping costs.
2. The system asks for payment method and the consumer chooses to pay through a credit card using SET.
3. Immediately, a special software on the consumer's PC called a Digital Wallet is invoked and it asks the customer to choose one credit card from the many he or she possesses.
4. The consumer chooses a card and the electronic transaction using SET is underway.
5. After getting details of customer payment the merchant contacts the merchant's Bank for customer authorization and payment.
6. Merchant Bank will contact the customer's Bank for the same and get approval of payment.
7. Merchant will notify, if transaction is successful.
8. A few seconds later, there is a confirmation to the customer that this order has been processed.
SET is a good example of a protocol that ignores the user authentication. SSL-based methods, on the other hand, are ignoring important "security" requirements [10]. Some disadvantages of SET are:
a) SET is designed for wired networks and does not meet all the challenges of wireless network.

b) As the SET protocol was designed to preserve the traditional flow of payment data (CA – MA – Merchant's Bank), an end-to-end security mechanism was required.
c) The third element is the direction of the transaction flow. In SET transactions are carried out between Customer Agent and Merchant. It is vulnerable to attacks like transaction/balance modification by Merchant.
d) The transaction flow is from Customer to Merchant so all the details of the users credit cards/debit cards must flow via the merchant's side. It increases the user's risk, since data can be copied and used later to access a customer account without authorization.
e) There is no notification to the Customer from the customer's Bank after the successful transfer. The user has to check his/her balance after logging on to his/her bank's website again.
f) SET is only for card (credit or debit) based transactions. Account based transactions are not included.

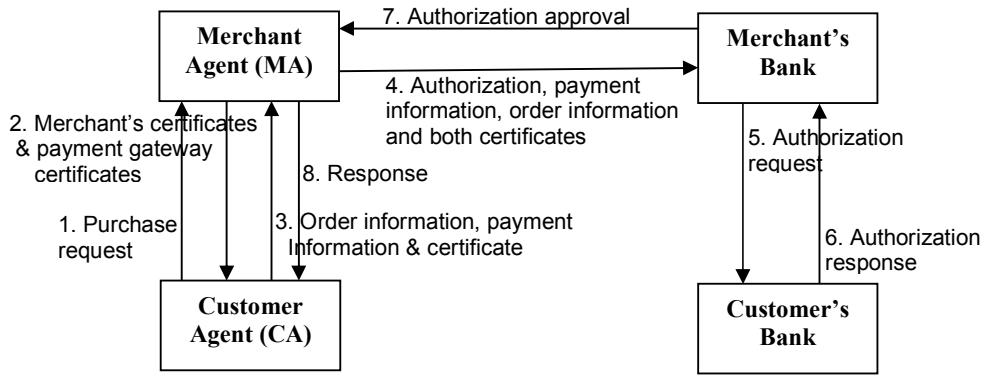

Figure 1. Transaction flow in Secure Electronic Transaction (SET)

## 3. MULTIFACTOR AUTHENTICATION APPROACH

Single-factor authentication, as in SET, is inadequate for high-risk transactions involving access to customer information or the movement of funds to other parties. Multifactor Authentication techniques can be used to provide secure web transactions using cell phones. Various methods have been proposed to provide multifactor authentication, like biometrics, extra hardware and software etc. However, the most practicable ones are those based on two separate communication channels. This technique is used in the present work.

## 3.1 MULTIFACTOR AUTHENTICATION TECHNIQUES

In the present work, we propose a multifactor authentication technique based on TIC's and SMS confirmation. *TIC Authentication:* TIC authentication is the technique which is used to verify both the user and the ongoing transaction. A TIC code certifies that the current transaction has been initiated by the right person and that it is a valid user who is trying to access his/her account. TIC codes are:
- Issued by the Bank or Financial institution to its customers.
- An 8 bit or 16 bit Pseudo randomly generated code which is assigned to the customers.
- May be a complicated digit sequence or combination of numeric and alpha numeric characters.

In our protocol the user will get a list of TIC codes from the bank or financial institution according to their requirement. TIC codes are encrypted and stored on user's cell phone. The key is a local password on the cell phone and user can change it easily. The Bank or Financial institution will keep record of issued TIC codes to its customers and match the same code during the online web transaction. A TIC code is cancelled after each transaction (i.e. each TIC code is used only once). Financial institutions can also decide to designate a validity time period of TICs according to its standard organizational issuing policies, providing an extra security feature.

*SMS Authentication:* Another method to validate user transaction is SMS confirmation. The financial institution stores user cell phone number to provide multi-factor authentication [1]. We assume that users carry their cell phone with them regularly and therefore can receive the short message. As a result, only valid users will receive an SMS from the authentication server. After getting the SMS, a user can acknowledge the choices. When the authentication server receives "YES" it knows that the user is valid and that the user has approved the initiated transaction. So, Multifactor Authentication is used to verify the user and the transaction by using following steps:

1. *Web-Based Basic Authentication*: Firstly, the user has to prove his identity to the Web server using a Web-based username/password based protocol for basic authentication.
2. *TIC Authentication*: After authenticating the web user using username/password, the Web server demands a TIC code from the web user. Now user will decrypt and insert one time TIC code to uniquely identify his/her transaction and prove his/her identity to the web Authentication server.
3. *SMS Confirmation*: After the TIC code identification, the web user will get an SMS from the web Authentication server to confirm his/her financial transaction. The security of the system also depends on the security of the messages sent by SMS and WAP, which are encrypted with A5/3 Algorithm [13]. By this SMS, a user can confirm his transaction by "YES" or "NO" [1]. Transaction is committed only if the user chooses "YES". The next Section gives the complete protocol proposed based on the above technique.

## 4. SECURE WEB AUTHENTICATION PROTOCOL

The data flow and architecture, based on Multifactor Authentication techniques, is described in this section.

## 4.1 ARCHITECTURE OF SECURE WEB AUTHENTICATION PROTOCOL

Figure 2 shows the protocol for a secure web authentication using a cell phone/PDA. This protocol starts when the user has decided to perform a money transfer operation. Here we assume that the user's cell phone number is stored on the authentication server. A separate authentication is recommended to maintain strong security.

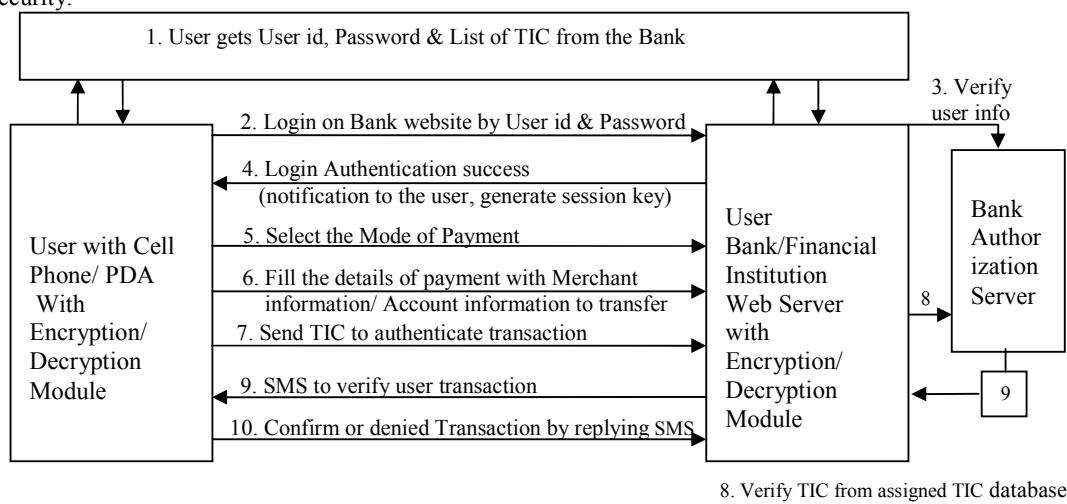

Figure 2: Protocol for wireless payment: Multifactor secure Web authentication protocol using mobile devices

As illustrated in Figure 2, following are the steps:
1. The user will login using a Username/Password and get a Transaction Identification code (TIC) from the bank. Each user has only one Username/Password to his/her account, but the TIC code is unique for each online transaction.
2. A Web-based username/password basic authentication is used to identify the user to the Web server.
3. After basic authentication the user will get an option to initiate transaction with a welcome message. We consider three modes of payment: Credit Card, Debit Card and Electronic transfer.
4. The user will get a notification of a successful logging with welcome message and display session key.
5. The user will select mode of payment.
6. The user will insert the details of payment by filling in a simple form with details such as the merchant's account number, invoice number or account number to which an amount has to be transferred.

7. The user will insert a TIC code by simply choosing a TIC code from the stored list of TIC codes. Note that TICs are password protected on the cell phone.

8. The bank authorization server verifies the TIC sent by the user by comparing it to its stored list of TIC's in the user account information at the server database. If both TIC's matched, it cancels the used TIC from its database and goes to the next step. If no TIC matched with the database then the authentication server will deny any further user transaction and display an appropriate error message to the user.

9. The authorization server will send an SMS to the user cell phone to verify his/her web transaction. The cell phone number of the user is available on the authorization server.

10. The user will confirm his/her initiated transaction by choosing "YES", or deny it by choosing "NO", by sending a confirmation SMS.

## 4.2 CRYPTOGRAPHY AND KEY MANAGEMENT

Public key encryption techniques have been adopted in many areas including network security, operating systems security, application data security and Digital Rights Management (DRM). Internet standardization bodies, such as the Internet Engineering Task Force (IETF) are constantly influencing the standardization process of the mobile platforms and specifically the cellular environment. Consequently, cellular-related standards have adopted Public Key Infrastructure (PKI) as a fundamental element in the construction of security for the near and far future of a wireless environment. [15]. Compared to symmetric-key encryption, public-key encryption needs more computation and is therefore not always appropriate for large amounts of data. However, it is possible to use public-key encryption to exchange a symmetric key, which can then be used to encrypt additional data. This approach is frequently used in many security protocols and is called *hybrid encryption schemes* [14, 15]. In this paper we have implemented a hybrid encryption scheme over wireless medium to get security. A TIC code is used to encrypt all the account details of the customer before submission to the server, and then the TIC code itself is encrypted with a secret key which is generated by the server and transmitted to the user after a successful login. After encryption, this TIC code together with transaction details will be transferred to the server. On the server side, an identical secret key is stored which will decrypt the TIC code, then match this TIC code with the TIC's issued to the customer. If this TIC matches with the database then it will be next used to decrypt the other transaction details which were encrypted by an identical TIC at the user end. At the start of every user session the server randomly generates a secret key and stores it in the user's specific entry in the database. It is a unique key for each user session. The server then encrypts this secret key using the client's 64-bit PIN code, which is a shared secret known to client and server. This encrypted secret key is transmitted to the client where it is stored in local variables. The client decrypts this secret key and uses it to encrypt the TIC code before transmission. However, there are several instances when we require a two-way authentication. In the following section we present the protocol for this purpose.

## 5. SYSTEM FOR TWO WAY AUTHENTICATION

After having analyzed the Secure Electronic Transaction (SET), on-line payments [6, 10] and having taken into consideration the constraints of the wireless infrastructure, we developed the secure protocol for Wireless Payment, supportive of one-way authentication in the previous section. We extend the protocol to support two-way authentication in the present section. In this architecture (Figure 3) we have considered five major components with certain roles: 1) *User*: A user is a valid account holding customer of the bank, 2) *Customer Agent (CA):* A CA is a software module which is running on the customer's mobile device, 3) *Merchant Agent (MA):* An MA is an online service provider and merchant website by which users do online purchasing, 4) *Customer's Bank*: This is the bank at which the user has a valid account, it also contains the authentication server necessary to authenticate the user, 5) *Merchant Bank*: This is the bank in which the merchant has a valid account, the merchant Bank is also responsible for authenticating the merchant. The two way Authentication protocol commences when the MA sends to CA an invoice detail and terminates when the MA receives a confirmation of payment from the Merchant's Bank. As shown in Figure 3, the flows for the payment transactions are as follows:

1. The MA prepares an invoice and sends the Merchant's encrypted banking information and authentication certificate with the invoice details to the CA.

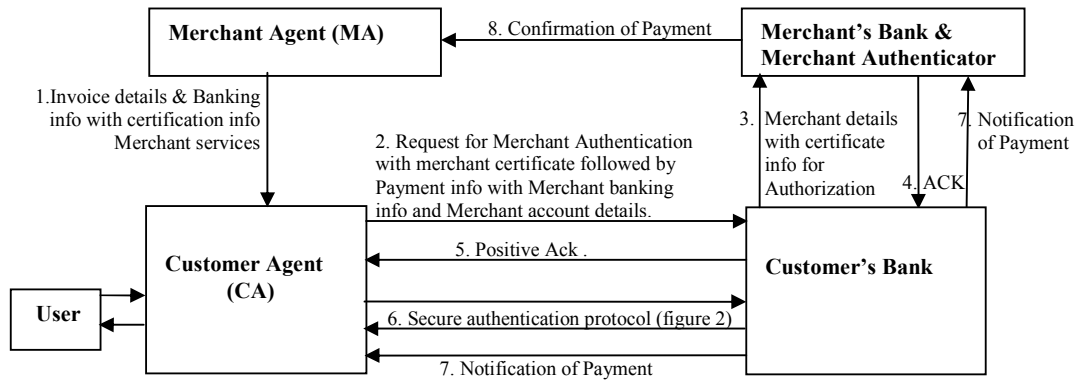

Figure 3: Protocol for Wireless Payment: Two Way Authentication system

2. The CA requests for authentication of Merchant to his Bank with the Merchant bank details, the merchant account details and a Merchant authentication certificate provided by the MA.
3. The Customer Bank forwards the Merchant details with the authentication certificate to the Merchant bank for authentication of the merchant.
4. The Merchant Bank sends a positive or a negative acknowledgement to the Customer Bank which confirms the validity of the Merchant or invalidates the Merchant.
5. In case of validation, the Customer's bank sends a Positive ACK to the CA and goes to step 6. If the Merchant certificate is not valid, the Customer Bank will notify the CA with that information. If Customer Bank received a negative or suspicious acknowledgement of the Merchant it simply rejects the user transaction with valid security reason.
6. Secure web authentication protocol will run (Figure 2) to authenticate the customer before payment after a valid authentication of the merchant. If the customer Bank received a positive Acknowledgement of the Merchant it demands a TIC for Customer Authentication. After aTIC verification it sends a confirmation SMS to the customer to verify his transaction using the secure web Authentication protocol (Figure 2).
7. After getting an SMS confirmation from the customer, the Customer Bank will send payment notification to the Merchant Bank as well as to the customer with the customer details and an invoice number and amount.
8. Merchant bank will send a confirmation of the received payment from the customer to the MA with relevant details, such as invoice number, customer id and amount received.

   We do not route payment transaction data via the MA. As a result, the implementation of security is less onerous. Customer payment instructions are no longer sent to the merchant and thus cannot be altered by the merchant. The two-way authentication protocol addresses several of the shortcomings of the SET. For example, data that is vital for the user is never available to the merchant in an unencrypted manner. Moreover, even if the cell phone is stolen and the user's password is available to the thief, the system is still secure, because the TIC's are also encrypted and they can be decrypted only by a key known to the valid user. The protocol is also secure against "man-in-the-middle" attacks since at no stage do we send unencrypted data over untrusted network. Also, we have the benefits of OTP since each TIC is used for only one transaction. The actual implementation requires elaboration of some specific technologies which are discussed in the next section.

## 6. IMPLEMENTATION ISSUES

J2ME is the preferred development platform due to the following reasons: the portability of Java code, the ability to reduce the network traffic by making use of the processing power of the Java phone to process data locally, and the ability to establish a differential security policy on the client that will perform the encryption operations according to the content and sensitivity of network data rather than encrypting everything. This helps us utilize the embedded device processing power very efficiently. Also, J2ME mobile information device applications (MIDlets) can operate over and make use of, the WAP stack to perform HTTP network interaction, without requiring TCP/IP. A side-effect of devising a general application-layer security solution using J2ME is that it provides a feasible solution to the traditional security gap in the WAP gateway [11]. This security gap is due to the security protocol conversion mechanism taking place in the WAP gateway between the secure sockets layer (SSL) encryption and the WAP wireless transport layer security (WTLS) encryption protocols. This protocol conversion leaves data in an unencrypted form at the gateway during the protocol switching process, which increases the risks to the confidentiality of data in the gateway [12].

HTTP is a stateless protocol. Each HTTP request is independent of the other requests and the protocol specification does not devise any mechanism that can group a series of requests as belonging to one user or another. Over time, several strategies have evolved to address session tracking; the most practical of these are the use of cookies and URL rewriting. The Java Servlet API provides the HttpSession object, which represents each user's interaction with the web server. The Servlet API specification requires that web "containers" implement session tracking using the cookie mechanism. The cookie interchange mechanism is similar to that taking place between a web browser and a web server with one difference – in our proposed protocol, the client MIDlet program has to explicitly send the session cookie back to the server for every request.

## 6.1 SIMULATION

Our simulation of client applications has used The Sun J2ME Wireless Toolkit consisting of build tools, utilities and a device emulator. It also includes the standard APIs like Limited Device Configuration (CLDC), Mobile Information Device Profile (MIDP), Wireless Messaging API (WMA), PDA Optional Packages for the J2ME Platform, J2ME Web Services Specification etc. The authentication server is based on J2EE technology with web server Apache Tomcat and database Oracle 9i with Jserver capabilities. The application starts with client performing HTTP request to activate server java servlet for login. It has the following tasks:
1. Authenticate user login and secret password and generate welcome message to the user.
2. Generate session key for new user session on successful login and store it on client specific entry in database at server side.
3. Generate secret key for user session after successful login and execute encryption method to encrypt these keys by using shared secret logic (discussed in section 4.2).
4. Transfer Keys to the user with welcome message and maintain session tracking for the session.

At client side, session key would be stored in cookies for life time of session and user variables are used to store secret key after decryption using shared secret logic (discussed in Section 4.2). Next the user will select mode of payment by executing another java servlet and enter the details of transaction by filling details like amount to be transferred, account number to which it has to be transferred, bank branch code, etc. In case of payment to any merchant it will automatically pickup the data like invoice number and merchant details which are given by merchant after selection of products and choosing payment option. At client side, the next step is to insert the TIC code in on going payment transaction using the menu option. User will insert local password to open the TICs list and pick any one of the TIC for transaction. The chosen TIC is automatically removed from the stored list of TIC's and displayed on the user screen. Next step is the encryption of sensitive data of transaction and TIC which we already discussed in Section 4.2. At server side two servlets would be responsible for calling the decryption methods to decrypt the TIC and Transaction details (as discussed in Section 4.2) and perform authentication.

Another important module in the protocol is generation of TIC codes at the financial institution server, distribution of TIC's to the customer and encryption of TIC's before storing them on client environment. TIC codes are pseudo random codes and can be generated with pseudo random number generation algorithm as mentioned in [18,19]. TIC generation logic is strictly confedentially at the web authentication server and we are assuming that the banks will update TIC generation data and improve TIC generation algorithms. Users demand TIC's as per their requirements as suggested in Sections 3.1 and 4.1. The authorized person of the financial institution is responsible for the distribution of TIC's to the user cell phone via simple data cable or a Bluetooth device and distribution process includes the encryption of TIC's for security reasons. At server side, we have assumed TIC's are stored in database and there is strong security of Database management system (supported by Oracle 9i) and operating system with secure firewalls to protect server side data. To satisfy need of TIC's of large number of users is part of future work. The implementation of this protocol will not increase expenses of users significantly. This protocol can be easily implemented and executed on the current expenses charged by financial institution from the users to perform online payments or with very less addition to the current charge of online payment. Basically cost model of the suggested protocol depends mostly on the policies that financial institutions adopt for implementing this protocol.

## 7. CONCLUSION

In this paper, we have presented an application-layer security solution for wireless payment system to provide end-to-end authentication and data confidentiality between wireless J2ME based clients and J2EE based servers. This paper suggests a new protocol for web user authentication based on multifactor authentication approach which is completely secure and easy to implement. We also suggest an approach for two-way authentication protocol to authenticate both the parties. This solution can be implemented within the limited resources of a Java MIDP device, without any modification to the underlying protocols or wireless network infrastructure. Future work will focus on developing a new and efficient way to get TIC codes from the financial institutions. TIC code installation on the user's cell phone must also be an easy task to avoid repeated visit by user to the bank or financial institution. Server side TIC maintenance and management mechanism to satisfy the demand from a large number of users is also part of future work.